\documentclass[proceedings, preprint]{rmaa}

\usepackage{paralist}

\usepackage{psfrag,color}
\usepackage[flushleft]{threeparttable}

\SetYear{2023}
\SetConfTitle{7th Workshop on Robotic Autonomous Observatories}

\title{The core collapse of a 16.5\,M$_{\odot}$ star} 
\author{
  Amar Aryan\altaffilmark{1,2}
  Shashi Bhushan Pandey,\altaffilmark{1}
  Rahul Gupta,\altaffilmark{1,3,4} 
  Amit Kumar Ror,\altaffilmark{1} 
  and A. J. Castro-Tirado\altaffilmark{5,6}}

\altaffiltext{1}{Aryabhatta Research Institute of Observational Sciences (ARIES), Manora Peak, Nainital-263002, India.}

\altaffiltext{2}{Graduate Institute of Astronomy, National Central University, 300 Jhongda Road, 32001 Jhongli, Taiwan}

\altaffiltext{3}{Astrophysics Science Division, NASA Goddard Space Flight Center, Mail Code 661, Greenbelt, MD 20771, USA}

\altaffiltext{4}{NASA Postdoctoral Program Fellow}

\altaffiltext{5}{Instituto de Astrof\'isica de Andaluc\'ia (IAA-CSIC), Glorieta de la Astronom\'ia s/n, E-18008, Granada, Spain.}

\altaffiltext{6}{Unidad Asociada al CSIC Departamento de Ingenier{\'i}a de Sistemas y Autom{\'a}tica, Escuela de Ingenier{\'i}as, Universidad de M{\'a}laga, C\/. Dr. Ortiz Ramos s/n, 29071 M{\'a}laga, Spain.}

\shortauthor{Aryan et al.}
\shorttitle{The core collapse of a 16.5\,M$_{\odot}$ star}

\listofauthors{Aryan et al.}

\abstract{We investigate the 1D stellar evolution of a 16.5\,M$_{\odot}$ zero-age main-sequence star having different initial rotations. Starting from the pre-main-sequence, the models evolve up to the onset of the core collapse stage. The collapse of such a massive star can result in several kinds of energetic transients, such as Gamma-Ray Bursts (GRBs), Supernovae, etc. Using the simulation parameters, we calculate their free-fall timescales when the models reach the stage of the onset of core collapse. Estimating the free-fall timescale is crucial for understanding the duration for which the central engine can be fueled, allowing us to compare the free-fall timescale with the T$_{\rm 90}$ duration of GRBs. Our results indicate that, given the constraints of the parameters and initial conditions in our models, rapidly rotating massive stars might serve as potential progenitors of Ultra-Long GRBs (T$_{\rm 90}$ $>>$ 500 s). In contrast, the non-rotating or slowly rotating models are more prone to explode as hydrogen-rich Type IIP-like core-collapse supernovae.}

\resumen{
Investigamos la evoluci\'on estelar 1D de una estrella de secuencia principal de edad cero de 16,5\,M$_{\odot}$ que tiene diferentes rotaciones iniciales. A partir de la secuencia previa a la principal, los modelos evolucionan hasta el inicio de la etapa de colapso del n\'ucleo. El colapso de una estrella tan masiva puede dar lugar a varios tipos de fuentes transitorias de alta energ\'ia, como explosiones de rayos gamma (GRB), supernovas, etc. Utilizando los par\'ametros de la simulaci\'on, calculamos sus escalas de tiempo de ca\'ida libre cuando los modelos alcanzan la etapa de inicio del colapso del n\'ucleo. Estimar la escala de tiempo de ca\'ida libre es crucial para comprender la duraci\'on durante la cual se puede alimentar el motor central, lo que nos permite comparar la escala de tiempo de ca\'ida libre con la duraci\'on T$_{\rm 90}$ de los GRB. Nuestros resultados indican que, dadas las limitaciones de los par\'ametros y las condiciones iniciales en nuestros modelos, las estrellas masivas que giran r\'apidamente podr\'ian servir como progenitores potenciales de GRB ultralargos (T$_{\rm 90}$ $>>$ 500 s). Por el contrario, los modelos en los cuales los progenitores no giran o que giran lentamente son m\'as propensos a que se produzcan supernovas de colapso del n\'ucleo de clase IIP que son ricas en hidr\'ogeno.
}

\addkeyword{Simulation tools: MESA}
\addkeyword{Simulation tools: STELLA}
\addkeyword{Stars: Zero-age Main-sequence}
\addkeyword{Stars: Collapsing}
\addkeyword{Transients: Gamma-Ray Bursts}
\addkeyword{Transients: Core-collapse Supernovae}

\begin{document}
\maketitle

\section{Introduction}
\label{sec:intro}

The core collapse of massive stars could result in several classes of transients. Usually, the collapse of massive stars having initial masses $\gtrsim$\,8\,M$_{\odot}$ produces the catastrophic transient phenomena known as core-collapse supernovae (CCSNe; for review, please see \citealt[][etc.]{1986ARA&A..24..205W,doi:10.1126/science.1100370,2005NatPh...1..147W}); however, the association of SN component with a few long gamma-ray bursts (LGRBs; for review on SN-GRB connection, please see \citealt[][etc.]{1999Natur.401..453B, 2006ARA&A..44..507W, 2016ApJ...832..108M}) have opened new avenues of transients resulting from the collapse of massive stars. Such identified connections of SN and GRB have indicated that a fraction of GRBs come from the collapse of massive stars \citep[][]{2017AdAst2017E...5C}. In recent years, observational technologies and theoretical modeling advancements have significantly enhanced our understanding of the evolution of massive stars as the progenitors of CCSNe and a special class of GRBs. The detection of SN progenitors in high-resolution pre-explosion images has enabled us to put important constraints over the evolution and terminating phases of massive stars \citep[][]{2015PASA...32...16S,2017RSPTA.37560277V}.  Besides the direct detection of progenitors in pre-explosion images, serious simulations have been conducted to constrain the properties of massive star evolution. In their simulations, \citet[][]{2018ApJ...859...48P} employed stars with the masses of 30 M$_{\odot}$ and 40 M$_{\odot}$ under various initial rotation conditions. Their research revealed that moderately rotating massive stars could conclude their evolution as BSG, capable of launching ultra-relativistic jets to power ULGRBs. Furthermore, \citet[][]{2023arXiv230105401S} extensively explored the effects of initial mass, metallicity, and rotation on magnetar formation, which could serve as the central engine to power the GRBs. In a very recent study, \citet[][]{2024arXiv240601220R} studied the evolution of massive stars having initial mass in the range of 15--30\,M$_{\odot}$ to constrain the properties of ULGRB progenitors.

The lower mass limit for typical LGRB progenitors, as suggested by \citet{2007MNRAS.376.1285L}, is close to 20\,M$_{\odot}$. However, modeling results from \citet[][]{2018ApJ...859...48P} identified BSG stars as ULGRB progenitors, with a standard mass of 15\,M$_{\odot}$ \citep[][]{2018arXiv181207620D}. Following these studies, the lower limit of mass for a progenitor to produce a GRB could lie in the range of 15--20\,M$_{\odot}$. Supported by this mass range, we choose a 16.5\,M$_{\odot}$ zero-age main-sequence (ZAMS) star and investigate its entire evolution on HR diagram. Based on the minimum mass limits to produce a CCSN or an LGRB, a star with an initial mass of 16.5\,M$_{\odot}$ capable of producing both, a CCSN or an LGRB, depending upon the initial rotation and metallicity. We are also slightly biased to choose a 16.5\,M$_{\odot}$ because one of our nearby Red supergiant (RSG) in the Orion constellation, Betelgeuse, has a similar mass \citep[][]{2020ApJ...902...63J}.

The entire manuscript is structured as follow; in \S~\ref{sec:intro}, we provide a broad overview of the current progress in the massive star evolutions and the motivation behind selecting a 16.5\,M$_{\odot}$ ZAMS star for the present study. In \S~\ref{sec:nums}, we provide the details of the numerical setups to evolve the various non-rotating and rotating models up to the core collapse stage. In \S~\ref{sec:prop}, we present a broad discussion on the physical properties of models under concern. \S~\ref{sec:GRBs} presents the details of the collapse of rapidly rotating models and the comparison of their properties with the actual observables of ULGRBs; additionally, the CCSNe resulting from the core-collapse of non/slowly rotating models are also discussed in this section. Finally, in \S~\ref{sec:results}, we provide a brief discussion and major outcomes of the present study.

\section{Core Collapse of the 16.5\,M$_{\odot}$ star: Numerical Setups}
\label{sec:nums}


In this section, we highlight the specific numerical setups for the simulation of a 16.5\,M$_{\odot}$ ZAMS mass star model having different initial rotations utilizing the state-of-the-art, 1D stellar evolution tool, the Modules for Experiments in Stellar Astrophysics {\tt MESA} of version 23.05. \citep{2011ApJS..192....3P, 2013ApJS..208....4P, 2015ApJS..220...15P, 2018ApJS..234...34P, 2019ApJS..243...10P, 2023ApJS..265...15J}. To evolve the models from their pre-main-sequence (PMS) stages up to the stage of the onset of core collapse, we utilize the {\tt 20M\_pre\_ms\_to\_core\_collapse} {\tt test\_suit} directory. We choose the initial conditions for the models considering various characteristics of exploding stars outlined in literature \citep{2016ApJS..227...22F,  2018ApJ...859...48P, 2018ApJS..234...34P, 2018ApJ...858..115A, 2023arXiv230105401S}. 
\begin{figure*}
\centering
    \includegraphics[height=7.0cm,width=8cm,angle=0]{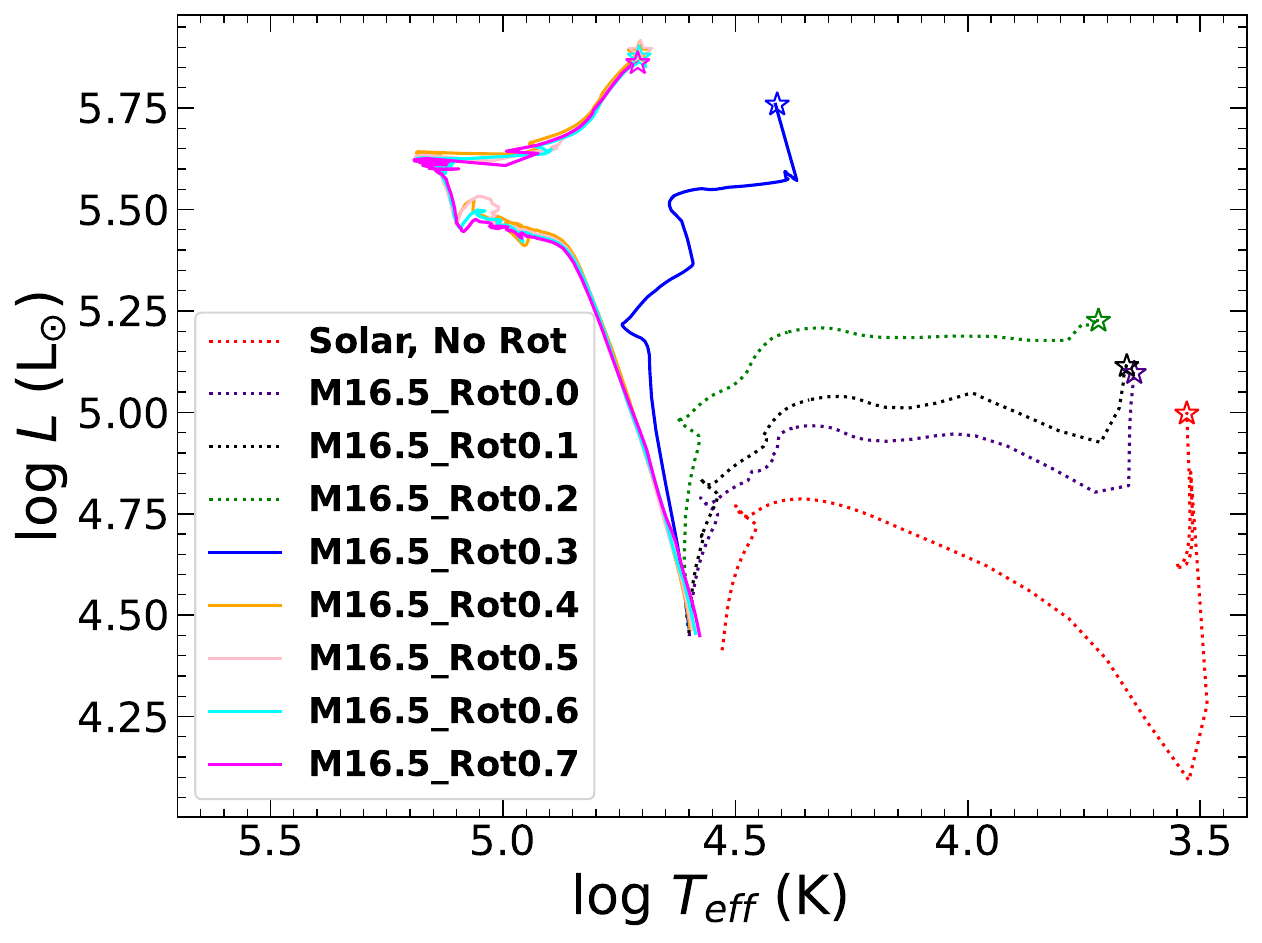}
    \includegraphics[height=7.0cm,width=8cm,angle=0]{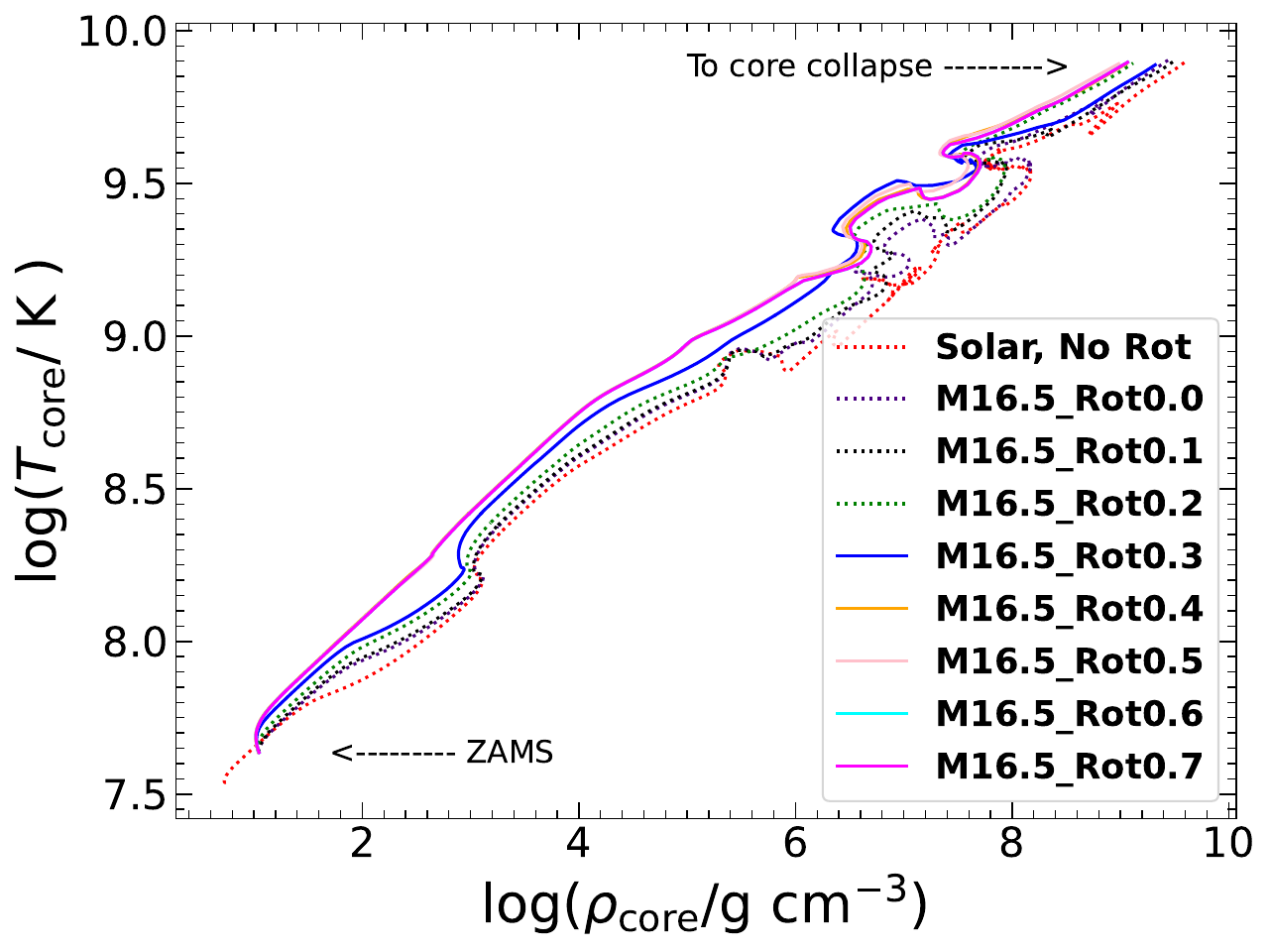}
   \caption{ {\em Left:} The evolution of our models on the HR diagram. The non/slow rotating models (Rot0.0, Rot0.1, and Rot0.2) have relatively smaller luminosity, and they are also relatively cooler compared to the faster rotating models when they are at the stage of the onset of core collapse (marked with $\star$). The non-rotating solar metallicity model is also shown for comparison purposes. { {\em Right:} The evolution of core-temperature and core-density, as the models evolve in time and reach the stage of the onset of core collapse.}}
    \label{fig:hr_rho}
\end{figure*}
\begin{figure*}
\centering
    \includegraphics[height=8.0cm,width=8cm,angle=0]{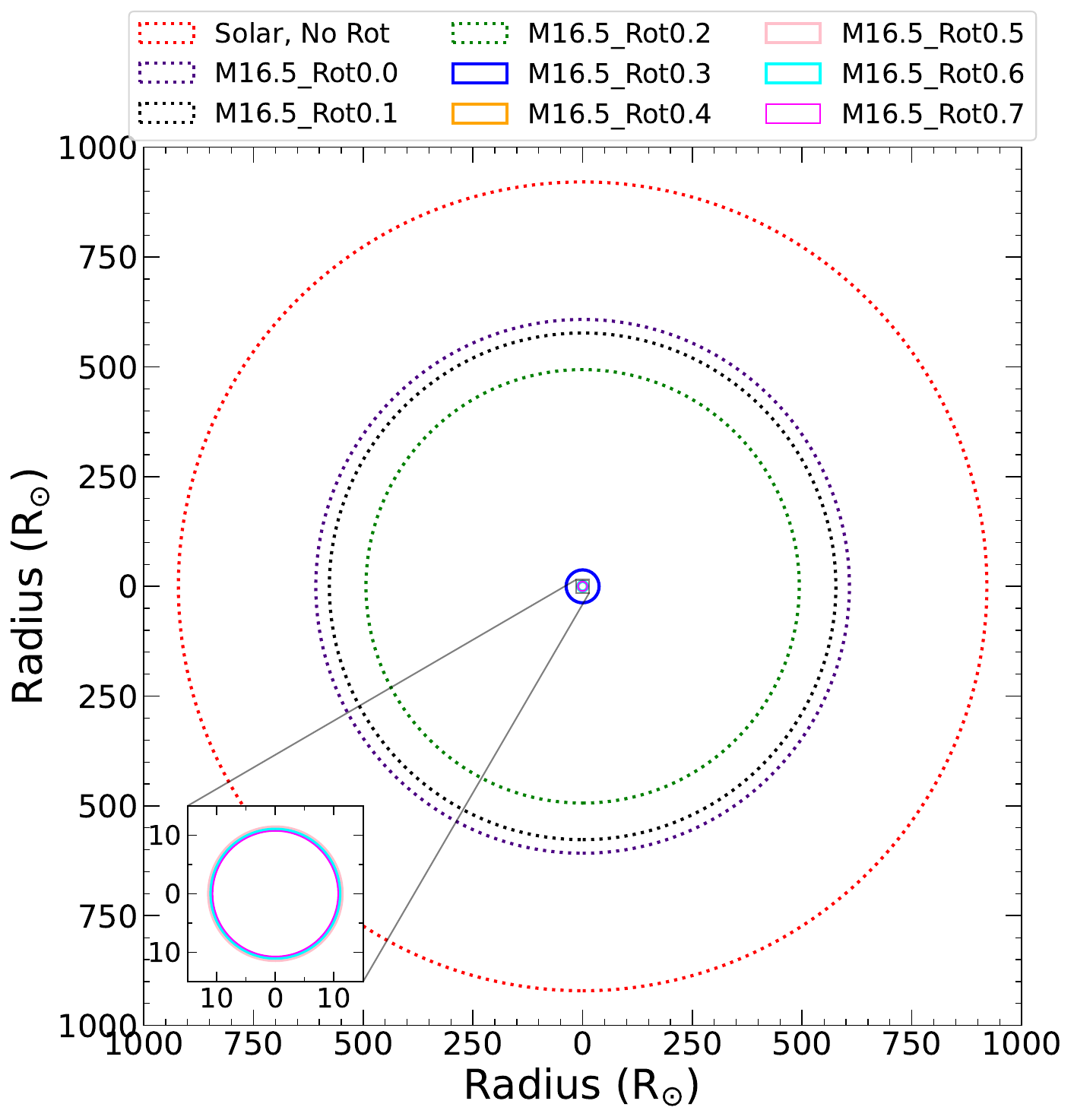}
    \includegraphics[height=8.0cm,width=8cm,angle=0]{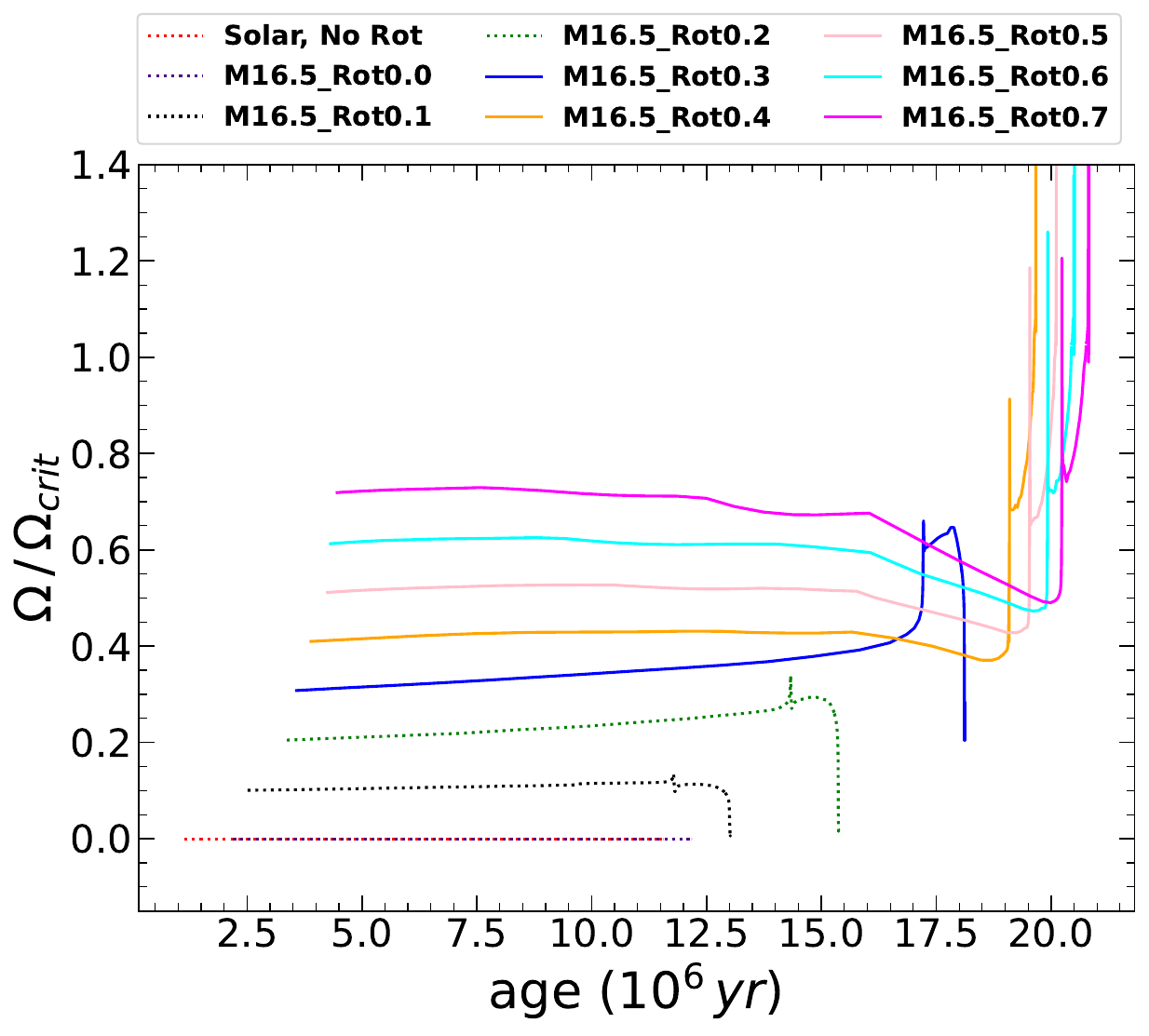}
   \caption{ {\em Left:} {The size of our models when they have arrived at the onset of core collapse. The rapidly rotating models are relatively more compact than the non/slow rotating models. {\em Right:} The variation of the rotational velocity throughout the course of the evolution of the models from the MS up to the stage of the onset of core collapse. The rapidly rotating models (except the model with Rot0.3) exceed their critical rotational velocity limits during the last stages of their evolution.}}
    \label{fig:radius_omega}
\end{figure*}
\begin{figure*}
\centering
    \includegraphics[height=8.0cm,width=8cm,angle=0]{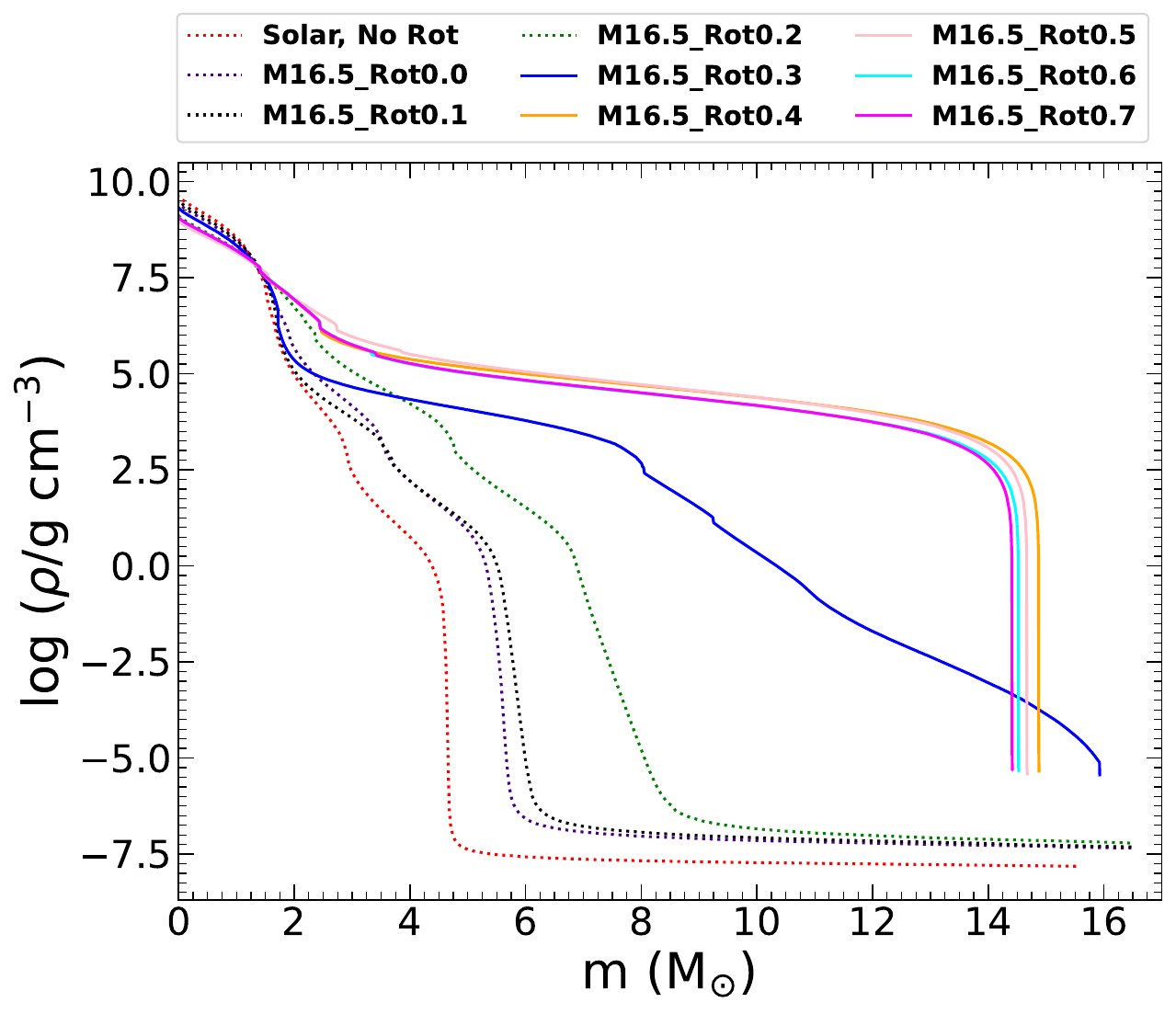}
    \includegraphics[height=8.0cm,width=8cm,angle=0]{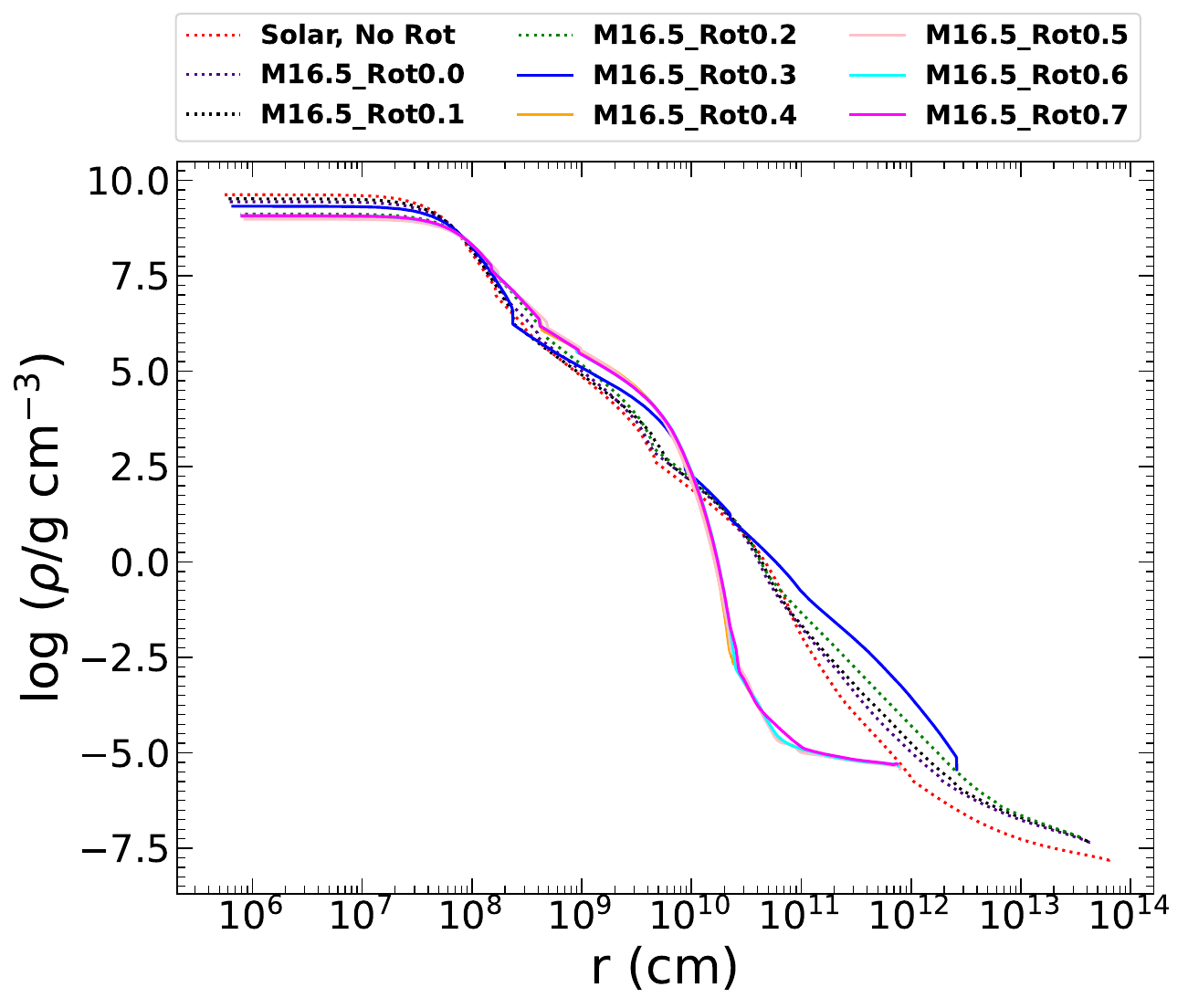}
   \caption{ {\em Left:} {The progenitor mass density profile of the models as a function of mass coordinate. Beyond the inner cores, the rapidly rotating models maintain a relatively higher density than the slow-rotating models. {\em Right:} The progenitor mass density profile of the models as a function of radius. The rapidly rotating models appear to maintain more compactness than the slow-rotating models; both of these profiles have been derived when all the models have reached the stage of the onset of core collapse. }}
    \label{fig:mass_rho_rad}
\end{figure*}

The variety of {\tt MESA} parameters in this study to evolve our models up to the onset of core collapse stage, closely follow the {\tt MESA} settings employed in our earlier studies (\citealt[][]{2021MNRAS.505.2530A, 2022JApA...43...87A, 2021MNRAS.507.1229P, 2022JApA...43....2A, 2022MNRAS.517.1750A, 2023arXiv230703234A, 2024arXiv240601220R}). However, we mention a few important modifications ahead. The stellar models in our study have the same ZAMS mass of 16.5\,M$_{\odot}$. We employ the metallicity (Z) of 2 $\times$ 10$^{-4}$ for each model (except the one with solar metallicity) in our study, which is favored by host galaxy observations of LGRBs \citep{2003A&A...400..499L, 2011MNRAS.414.1263M, 2022JApA...43...82G}. Starting from the non-rotating model, we change the initial angular rotational velocity ($\Omega$) in steps of 0.1\,$\Omega_{\rm c}$ till  0.7\,$\Omega_{\rm c}$.  Here, $\Omega_{\rm c}$ is the critical angular rotational velocity expressed as:

\begin{equation}
{\rm \Omega_{\rm c}^2} = (1-\frac{L}{L_{\rm edd}}) \frac{GM}{R^3} 
\label{eq:1}
\end{equation}

with $L_{\rm edd}$ representing the Eddington luminosity in the above equation (please see \citealt[][]{2011ApJS..192....3P, 2013ApJS..208....4P} for details on how {\tt MESA} implements rotation).

We apply the Ledoux criterion and model convection using the mixing length theory introduced by \citet[][]{1965ApJ...142..841H}, with a fixed mixing length parameter, $\alpha_{\rm MLT}$ of 2.0. To account for semiconvection, we set the semiconvection coefficient, $\alpha_{\rm sc}$ to 0.01, following the approach of \citet[][]{1985A&A...145..179L}. Thermohaline mixing is incorporated based on the formulation given by \citet[][]{1980A&A....91..175K}, using the efficiency parameter, $\alpha_{\rm th}$ of 2.0 till the stage of the core-He exhaustion and 0 thereafter, as employed in the default {\tt 20M\_pre\_ms\_to\_core\_collapse} {\tt test\_suit} directory of {\tt MESA}. Convective overshooting is modeled using parameters, $f_{\rm ov}$ = 0.005 and $f_{\rm 0}$ = 0.001, following \citet[][]{2000A&A...360..952H}, consistent with prior studies like \citet[][]{2016ApJS..227...22F} and \citet[][]{2023MNRAS.521L..17A}. The `Dutch' wind scheme, with a wind scaling factor ($\eta_{\rm wind}$) of 0.5, is employed to incorporate wind effects, aligning with previous works such as \citet{2018ApJ...859...48P}, \citet[][]{2018ApJ...858..115A}, and \citet[][]{2023arXiv230105401S}. A summary of some initial parameters can be found in Table~\ref{tab:mesa_table}.

\section{Core Collapse of the 16.5\,M$_{\odot}$ star: Physical properties}
\label{sec:prop}

With the above-mentioned {\tt MESA} settings of initial parameters, we evolve all the models from PMS up to the stage of the onset of core collapse. Following the default {\tt MESA} settings, the arrival of a model on ZAMS is marked at a stage when the ratio of the luminosity from nuclear reactions and the overall luminosity of the model becomes 0.4. Further, the beginning of the core collapse of the model is marked when the infall velocity of its Iron-core exceeds a limit of 100\,km\,s$^{-1}$. The left panel of Figure~\ref{fig:hr_rho} illustrates the evolutionary trajectory of the models in the current study on the HR diagram. Owing to the low initial metallicity, rotation, and a moderate wind scaling factor ($\eta_{\rm wind}$=0.5), the rapidly rotating models ( with $\Omega$\, $\ge$\,0.3\,$\Omega_{\rm c}$) terminate their evolution towards the relatively hotter end on the HR diagram in comparison to the slowly rotating models. 
The final temperatures of rapidly rotating models and their corresponding positions on HR diagram are more consistent with Blue supergiant (BSG) stars ($\Omega$\, $=$\,0.3\,$\Omega_{\rm c}$) or stripped Wolf-Rayet (WR) stars (models with $\Omega$\, $>$\,0.3\,$\Omega_{\rm c}$). On the other hand, the positions of the slowly rotating models on HR diagram are consistent with the locations of RSG stars ($\Omega$\, $<$\,0.3\,$\Omega_{\rm c}$). The right panel of Figure~\ref{fig:hr_rho} shows the variation of the core-temperature, log($T_{\rm core}$/K) and the core-density, log($\rho_{\rm core}$/g cm$^{-3}$) as the models pass through the various stages of their evolution starting from main-sequence (MS) upto the onset of core collapse. During the last evolutionary stages, the models seem to exceed the log($T_{\rm core}$/K) of 9.9 and corresponding log($\rho_{\rm core}$/g cm$^{-3}$) of 9. Such high temperatures and densities of the cores are considered ideal conditions for stars to collapse under their own gravity.  

\begin{table*}
\begin{threeparttable}
\small
\centering
\caption{Summary of the initial and final parameters employed in and derived from {\tt MESA} code} 
\begin{tabular}{|c|c|c|c|c|c|c|c|c|} \hline
\boldmath $\Omega/\Omega_{c}$ ~~ & ~~ \bf \boldmath M$_{\rm final}$ ~~ & ~~ \bf \boldmath R$_{\rm final}$ ~~ & ~~ \bf \boldmath M$_{\rm Fe-core}$ ~~ & ~~ \bf \boldmath log(T$_{\rm eff}$) ~~ & ~~ \bf Log (L) ~~ & ~~~~~~ \bf \boldmath t$_{\rm ff} ~~~~~ $ & ~~~~ \bf t$_{\rm b}$ ~~~~ \\ 

  & \bf \boldmath (M$_{\odot}$) & \bf \boldmath(R$_{\odot}$) & \bf \boldmath (M$_{\odot}$) & \bf \boldmath (K) & \bf \boldmath (L$_\odot$) & \bf (s) & \bf (s)\\ \hline

0.0 & 16.474 & 607.8 & 1.539 & 3.644 & 5.097 & 1385926.3 & 141.8\\
0.1 & 16.475 & 577.0 & 1.490 & 3.659 & 5.115 & 1281711.9 & 134.6\\
0.2 & 16.452 & 493.6 & 1.733 & 3.721 & 5.226 & 1014838.8 & 115.1\\
0.3 & 15.928 & 37.6 & 1.501 & 4.414 & 5.759 & 21663.6 & 8.8\\
0.4 & 14.878 & 11.2 & 1.937 & 4.708 & 5.885 & 3636.7 & 2.6\\
0.5 & 14.674 & 11.4 & 1.923 & 4.705 & 5.890 & 3773.7 & 2.7\\
0.6 & 14.528 & 11.0 & 1.840 & 4.709 & 5.875 & 3600.7 & 2.6\\
0.7 & 14.418 & 10.7 & 1.899 & 4.712 & 5.865 & 3477.0 & 2.5\\
\hline
\hline
Solar, No Rot  \\
\hline
\hline

0.0 & 15.519 & 921.0 & 1.520 & 3.529 & 4.998 & 2663287.0 & 214.8\\


\hline
\end{tabular}
\label{tab:mesa_table}
\begin{tablenotes}
      \small
      \item Note: Starting from the PMS phase, these models evolve until they arrive at the onset of core collapse stage. Here, M$_{\rm ZAMS}$ represents the model's mass at ZAMS, while $\Omega/\Omega{_c}$ is the ratio between initial angular rotational velocity and critical angular rotational velocity at ZAMS. Moreover, M$_{\rm final}$ denotes the final mass, R$_{\rm final}$ indicates the final radius, M$_{\rm Fe-core}$ represents the mass of the iron core, T$_{\rm eff}$ denotes the effective temperature, and L stands for the corresponding luminosity of the model at the onset of core collapse. Additionally, t$_{\rm ff}$ and t$_{\rm b}$ refer to the free-fall time of the model and the bore time of the weak jet, respectively.
    \end{tablenotes}
  \end{threeparttable}
\end{table*}

The left panel of Figure~\ref{fig:radius_omega} displays the final radii of our models at the stage of the onset of core collapse. The rapidly rotating models are quite compact having final radii of 10\,R$_{\odot}$ (models with $\Omega$\, $>$\,0.3\,$\Omega_{\rm c}$) or a few 10\,R$_{\odot}$ (the model with $\Omega$\, $=$\,0.3\,$\Omega_{\rm c}$). On the other hand, the non/slowly rotating models are enormous, having final radii of several 100\,R$_{\odot}$. The non-rotating solar metallicity model is much bigger than the low metallicity models. For a given mass and other simulation parameters, the radius decreases with a corresponding decline in metallicity due to the competing effects of compressional heating and radiative cooling (for details, please see \citealt[][]{2009ApJ...703.2205K}). These final sizes of the rapidly rotating and non/slowly rotating models are seemingly governed by the corresponding mass losses they suffer during their late evolutionary stages. As evident from the right panel of Figure~\ref{fig:radius_omega}, the rapidly rotating models either exceed their critical angular rotational velocities (model with $\Omega$\, $>$\,0.3\,$\Omega_{\rm c}$) or their angular rotational velocities become much higher than initial values (model with $\Omega$\, $=$\,0.3\,$\Omega_{\rm c}$). As a result, they suffer enormous mass loss, which almost completely strips off their outer hydrogen envelope, or a significant amount of their outer hydrogen envelope is lost \citep[][]{2023MNRAS.521L..17A,2024arXiv240601220R}. As a result, the corresponding final models are relatively much more compact when compared to the non/slowly rotating models, which have their outer hydrogen envelope (almost) intact.

Figure~\ref{fig:mass_rho_rad} illustrates the density profiles of the models as a function of their mass coordinates (left panel) and their radii (right panel) at the stage of the onset of core collapse. As seen in the left panel, the rapidly rotating models ($\Omega$\, $\ge$\,0.3\,$\Omega_{\rm c}$) have slightly smaller densities near the core when compared to non/slowly rotating models, however, as we move towards the surface, the rapidly rotating models maintain an overall higher density profile than the non/slowly rotating models. The reason for the non/slowly rotating models having an overall shallow density profile compared to rapidly rotating models can be attributed to their large hydrogen envelope. Moreover, the right panel shows that the rapidly rotating models have relatively higher densities and smaller final radii, which in turn implies that they are compact.

\section{Core Collapse of the 16.5\,M$_{\odot}$ star}
\label{sec:GRBs}
\begin{figure*}
\centering
    \includegraphics[height=7.0cm,width=8cm,angle=0]{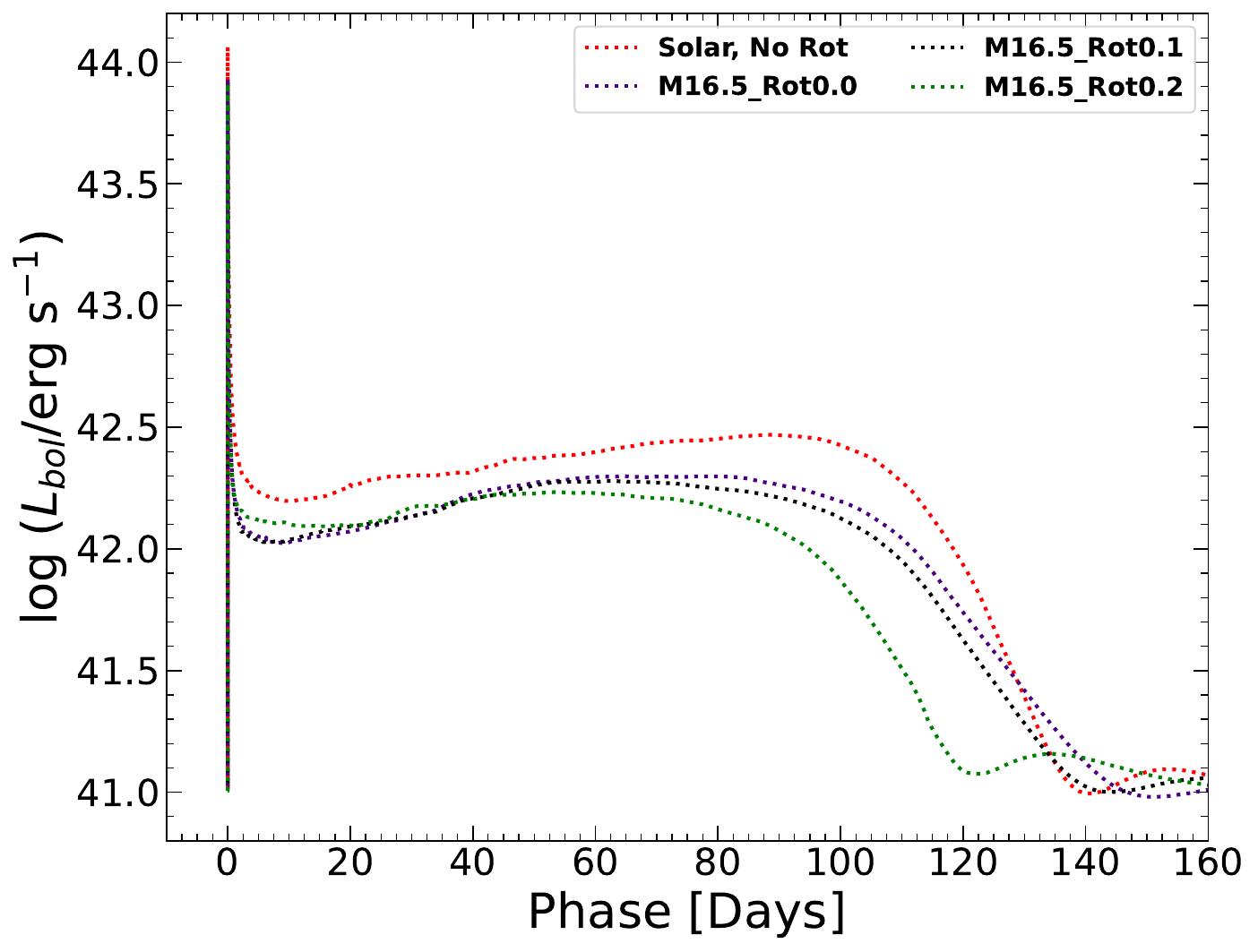}
    \includegraphics[height=7.0cm,width=8cm,angle=0]{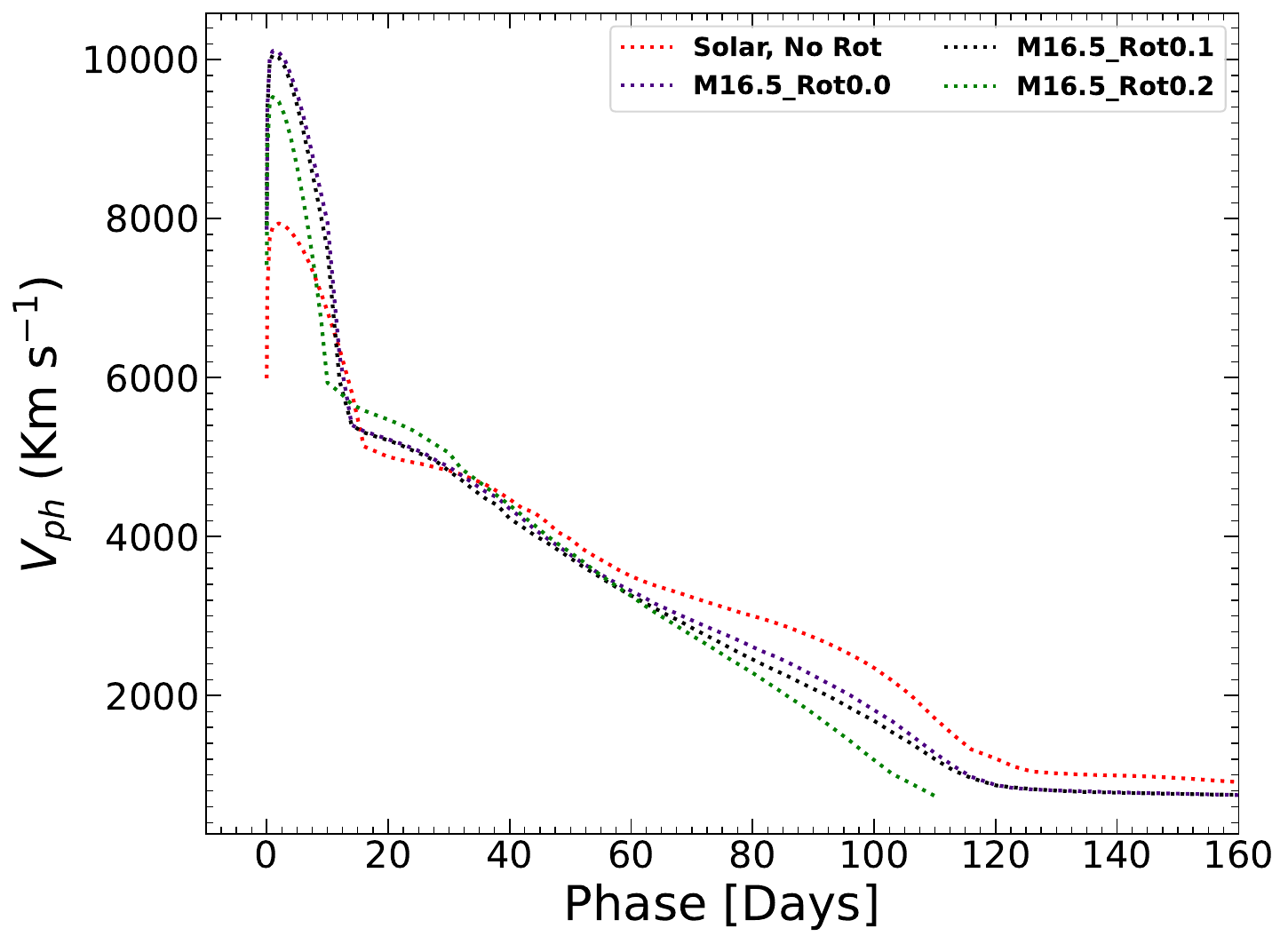}
   \caption{ {\em Left:} {The bolometric light curve of of CCSNe resulting from the explosion of slow rotating models. These light curves are identical to the Type IIP SNe light curve. {\em Right:} The corresponding photospheric velocity evolution. Once again, the photospheric velocity evolutions follow the trends of Type IIP SNe. The early and plateau phases luminosities of such SNe are well suitable to be observed by the BOOTES network of telescopes \citep[][]{1999A&AS..138..583C, 2023FrASS..10.2887H}.}}
    \label{fig:blc_vel}
\end{figure*}
\subsection{As the Progenitors of Ultra-long GRBs}
Using the simulation parameters of the models at the onset of core collapse, we calculate the free fall timescales (t$_{\rm ff}$) by applying Equation 1 from \citet[][]{2018ApJ...859...48P}. The t$_{\rm ff}$ for each model is detailed in Table~\ref{tab:mesa_table}. Estimating t$_{\rm ff}$ is crucial for understanding how long the central engine can be powered, enabling its comparison with the T$_{90}$ duration of GRBs. In a recent study by \citet[][]{2023ApJ...957...31S}, a two-stage model for GRB 221009A is proposed, linking the precursor pulse with the weak jet resulting from the core collapse of massive stars. Therefore, we also calculate the bore-time (t$_{\rm b}$) of the weak jet for each of the models to estimate the duration of the precursor pulse. The t$_{\rm b}$ for each model is estimated utilizing a simple equation given in \citet[][]{2024arXiv240601220R}:
\begin{equation}
{\rm t_{\rm b}} = \frac{\rm R_{\rm final}}{({\rm u}\Gamma)}
\label{eq:3}
\end{equation}
In the equation above, u represents the weak jet velocity corresponding to a Lorentz factor of $\Gamma$. Including $\Gamma$ in the denominator accounts for relativistic length contraction. When calculating t$_{\rm b}$ using Equation~\ref{eq:3}, we simplify by assuming the weak jet moves with a constant $\Gamma$ of 10. This choice is also supported by the analysis of \citet[][]{2023ApJ...957...31S} and \citet[][]{2024arXiv240601220R}. With these assumptions, the estimated t$_{\rm b}$ for each model is presented in Table~\ref{tab:mesa_table}. 

Now, we compare the t$_{\rm ff}$ derived from our simulation parameters with the T$_{\rm 90}$ duration of a few actual ULGRBs selected out of the Gold sample listed in Table A1 of \citet[][]{2024arXiv240601220R}. The {\tt M\_16.5\_Rot0.3} model (having 16.5\,M$_{\odot}$ of ZAMS mass and $\Omega$\, $=$\,0.3\,$\Omega_{\rm c}$) exhibits a t$_{\rm ff}$ of approximately 21600\,s, which closely aligns with the T$_{\rm 90}$ duration of GRB 111209A (T$_{\rm 90}$\,$\sim$\,18200s). Similarly, the {\tt M\_16.5\_Rot0.5} model displays a t$_{\rm ff}$ of around 3800\,s, close to the T$_{\rm 90}$ duration of GRB 090404 (T$_{\rm 90}$\,$\sim$\,4500s). Furthermore, the t$_{\rm ff}$ values obtained from the rapidly rotating models in our current study are of a comparable order when contrasted with the actual T$_{\rm 90}$ duration of the Gold sample of ULGRBs presented in \citet[][]{2024arXiv240601220R}.

As discussed in earlier sections, the non/slowly rotating models have large final radii (R$_{\rm final}$) at their terminating stage, moreover, their positions on HR diagram are consistent with massive RSG stars. The models terminating their evolution as RSGs cease to serve as the progenitors for the GRBs/ULGRBs since their enormous final radii ( $>$ several 100\,R$_{\odot}$, Table~\ref{tab:mesa_table}) do not allow successful penetration of the jet. Rather, the non/slowly rotating models are more prone to explode as CCSNe. In the next section, we investigate the properties when the non/slowly rotating models explode as CCSNe.

\subsection{As the Progenitors of CCSNe}

As mentioned, the parameters of non/slowly rotating models conflict for them to be the progenitors of ULGRBs; however, their core collapse can result in catastrophic CCSNe. To simulate their synthetic explosions for generating the bolometric light curves and corresponding photospheric velocity ($V_{ph}$) evolutions, we utilize a combination of {\tt MESA} and  {\tt STELLA} \citep[][]{1998ApJ...496..454B,2004Ap&SS.290...13B,2005AstL...31..429B,2006A&A...453..229B} following the process described by \citet[][in section 6]{2018ApJS..234...34P}. We utilize a nickel mass of 0.032\,M$_{\odot}$ and an explosion energy of 1.0$\times$10$^{51}$ for all the non/slowly rotating models. The choice of considered nickel mass and explosion energy are backed by \citet[][for nickel mass]{2019A&A...628A...7A} and \citet[][for explosion energy]{2003MNRAS.346...97N}.

The left panel of Figure~\ref{fig:blc_vel} shows the light curves of CCSNe resulting from the explosion of non/slowly rotating models. The light curves show plateaus similar to the Type IIP CCSNe. The right panel of Figure~\ref{fig:blc_vel} shows the corresponding photospheric velocity evolution of the models. Since these models have retained almost all of their outer hydrogen envelope, their light curves and photospheric velocity evolution are similar to Type IIP CCSNe.

\section{Discussion and Conclusions}
\label{sec:results}
In this work, we investigated the evolution of a 16.5\,M$_{\odot}$ ZAMS star with different initial rotations. Starting from the PMS, the models evolved until they arrived at the onset of core collapse. The physical properties of the models were then utilized to constrain whether they could serve as the progenitors of ULGRBs or CCSNe. Enumerated below were the findings from our analysis:
\begin{enumerate}
    \item The rapidly rotating models suffered significantly higher mass loss than the non/slowly rotating models. We noticed that the angular rotational velocities of rapidly rotating models exceeded their corresponding critical angular rotational velocities, resulting in enormous mass loss during the late evolutionary stages.
    
    \item The physical properties of rapidly rotating models were consistent with the BSG and WR stars. The comparison of their t$_{\rm ff}$ with the T$_{\rm 90}$ durations of ULGRBs mentioned in \citet[][]{2024arXiv240601220R} showed close resemblances. As an example, the t$_{\rm ff}$ duration for {\tt M16.5\_Rot0.5} was found to be close to the T$_{\rm 90}$ duration of GRB 111209A. The rest of the rapidly rotating models also showed the t$_{\rm ff}$ timescales having similar orders as the T$_{\rm 90}$ durations of several ULGRBs. Thus, the rapidly rotating models could serve as the progenitors of ULGRBs. 
    
    \item The physical properties of non/slowly rotating models were consistent with RGS stars. Their final radii were too high for them to be ULGRB progenitors. Rather, these non/slowly rotating models could explode as CCSNe, and the resulting light curves resembled the light curves of hydrogen-rich Type IIP SNe.

\end{enumerate}

\section*{Acknowledgements}
AA acknowledges funds and assistance provided by the Council of Scientific \& Industrial Research (CSIR), India, under file no. 09/948(0003)/2020-EMR-I. AA also acknowledges the Yushan Fellow Program by the Ministry of Education, Taiwan for the financial support (MOE-111-YSFMS-0008-001-P1). RG and SBP acknowledge the financial support of ISRO under AstroSat archival Data utilization program (DS$\_$2B-13013(2)/1/2021-Sec.2). AA, SBP, RG and AKR acknowledge support from DST grant no. DST/ICD/BRICS/Call-5/CoNMuTraMO/2023(G) for the present work. AJCT acknowledges support from the Spanish Ministry projects PID2020-118491GB-I00 and PID2023-151905OB-I00 and Junta de Andalucia grant P20\_010168 and from the Severo Ochoa grant CEX2021-001131-S funded by MCIN/AEI/ 10.13039/501100011033. RG was sponsored by the National Aeronautics and Space Administration (NASA) through a contract with ORAU. The views and conclusions contained in this document are those of the authors and should not be interpreted as representing the official policies, either expressed or implied, of the National Aeronautics and Space Administration (NASA) or the U.S. Government. The U.S. Government is authorized to reproduce and distribute reprints for Government purposes, notwithstanding any copyright notation herein.
\label{sec:refs}


\bibliography{references}

\end{document}